# Evolving the Digital Industrial Infrastructure for Production: Steps Taken and the Road Ahead

Jan Pennekamp, Anastasiia Belova, Thomas Bergs,
Matthias Bodenbenner, Andreas Bührig-Polaczek,
Markus Dahlmanns, Ike Kunze, Moritz Kröger, Sandra Geisler,
Martin Henze, Daniel Lütticke, Benjamin Montavon,
Philipp Niemietz, Lucia Ortjohann, Maximilian Rudack,
Robert H. Schmitt, Uwe Vroomen, Klaus Wehrle, and Michael Zeng

## Contents

J. Pennekamp (✉) · M. Dahlmanns · I. Kunze · K. Wehrle
Communication and Distributed Systems, RWTH Aachen University, Aachen, Germany
e-mail: pennekamp@comsys.rwth-aachen.de; dahlmanns@comsys.rwth-aachen.de;
kunze@comsys.rwth-aachen.de; wehrle@comsys.rwth-aachen.de

A. Belova · M. Bodenbenner · B. Montavon · P. Niemietz · L. Ortjohann
Laboratory for Machine Tools and Production Engineering, RWTH Aachen University, Aachen, Germany
e-mail: a.belova@wzl.rwth-aachen.de; m.bodenbenner@wzl.rwth-aachen.de;
b.montavon@wzl.rwth-aachen.de; p.niemietz@wzl.rwth-aachen.de;
l.ortjohann@wzl.rwth-aachen.de

T. Bergs
Laboratory for Machine Tools and Production Engineering, RWTH Aachen University, Aachen, Germany

Fraunhofer IPT, Aachen, Germany
e-mail: t.bergs@wzl.rwth-aachen.de

A. Bührig-Polaczek · M. Rudack · U. Vroomen
Foundry Institute, RWTH Aachen University, Aachen, Germany
e-mail: office.buehrig-polaczek@gi.rwth-aachen.de; m.rudack@gi.rwth-aachen.de;
u.vroomen@gi.rwth-aachen.de

**Abstract**

The Internet of Production (IoP) leverages concepts such as digital shadows, data lakes, and a World Wide Lab (WWL) to advance today's production. Consequently, it requires a technical infrastructure that can support the agile deployment of these concepts and corresponding high-level applications, which, e.g., demand the processing of massive data in motion and at rest. As such, key research aspects are the support for low-latency control loops, concepts on scalable data stream processing, deployable information security, and semantically rich and efficient long-term storage. In particular, such an infrastructure cannot continue to be limited to machines and sensors, but additionally needs to encompass networked environments: production cells, edge computing, and location-independent cloud infrastructures. Finally, in light of the envisioned WWL, i.e., the interconnection of production sites, the technical infrastructure

M. Kröger
Laser Technology, RWTH Aachen University, Aachen, Germany
e-mail: moritz.kroeger@llt.rwth-aachen.de

S. Geisler
Data Stream Management and Analysis, RWTH Aachen University, Aachen, Germany
e-mail: geisler@cs.rwth-aachen.de

M. Henze
Security and Privacy in Industrial Cooperation, RWTH Aachen University, Aachen, Germany
e-mail: henze@cs.rwth-aachen.de

D. Lütticke · M. Zeng
Information Management in Mechanical Engineering, RWTH Aachen University, Aachen, Germany
e-mail: daniel.luetticke@ima.rwth-aachen.de; michael.zeng@ima.rwth-aachen.de

R. H. Schmitt
Laboratory for Machine Tools and Production Engineering, RWTH Aachen University, Aachen, Germany

Fraunhofer IPT, Aachen, Germany

Information Management in Mechanical Engineering, RWTH Aachen University, Aachen, Germany
e-mail: r.schmitt@wzl.rwth-aachen.de

must be advanced to support secure and privacy-preserving industrial collaboration. To evolve today's production sites and lay the infrastructural foundation for the IoP, we identify five broad streams of research: (1) adapting data and stream processing to heterogeneous data from distributed sources, (2) ensuring data interoperability between systems and production sites, (3) exchanging and sharing data with different stakeholders, (4) network security approaches addressing the risks of increasing interconnectivity, and (5) security architectures to enable secure and privacy-preserving industrial collaboration. With our research, we evolve the underlying infrastructure from isolated, sparsely networked production sites toward an architecture that supports high-level applications and sophisticated digital shadows while facilitating the transition toward a WWL.

## 1 Introduction

With the deep integration of distributed, heterogeneous data producers and the incorporation of intelligent, reactive consumers that reliably exchange and evaluate data and make decisions in real time, the Internet of Production (IoP) is changing the requirements for the underlying physical information infrastructure. Concepts, such as digital shadows, data lakes of production, and the World Wide Lab (WWL) with its global knowledge exchange (Brauner et al. 2022), require a foundation that enables the seamless execution and transfer of physical, simulated, and data-driven production models and data streams with excessive peak loads in real time (Pennekamp et al. 2019a). The weakened boundaries between data processing and network communication and the gradual shift of computational tasks closer to the machines form the basis for the integration of complex, high-quality control with maximum flexibility into a decentralized infrastructure to enable the offloading of data-intensive tasks (Chang et al. 2014). Dynamic reconfigurability of the underlying architecture guarantees constant adaptation of the processes to the needs of production technology. Since a significant part of the value creation of the IoP is generated by the exchange of information between stakeholders from different, possibly mutually distrusting cooperation partners, an infrastructure must take confidentiality into account (Gelhaar et al. 2021). Despite extensive digitization, networking, and autonomy of production sites, humans also remain an important factor in the operation, maintenance, and optimization of plants, systems, and processes, as well as in decisions derived from an exchange of information (Neumann et al. 2021).

An underlying technical infrastructure that meets these requirements has to include all components of production sites, ranging from sensors and actuators that are integrated into production machines to distributed data centers in the cloud, as shown in Fig. 1. Digital shadows are a core concept of the IoP (Brauner et al. 2022) and correspond to representations of data that need to be handled within the infrastructure throughout all common states of data, i.e., at rest, in motion, and in use. Addressing the needs of the proposed WWL, the infrastructure further needs to be able to move data across stakeholders. In the underlying technical infrastructure, any processed data (and information) is thus part of constantly evolving digital shadows as the prevalent tasks of preprocessing, aggregation, filtering, data translation, and data sharing all constitute task- and context-dependent, purpose-driven, aggregated, and persistent representations and transformations of the originally sensed data. Building upon digital shadows that are primarily concerned with data processing and network communication within the technical infrastructure, we further require an infrastructure that is capable of effectively utilizing the information contained in digital shadows for decision-making. In light of this requirement, the infrastructure also needs to evolve the formal modeling of information (cf. chapter ▶ "The Internet of Production Digital Shadow Reference Model for Worldwide Production Labs"), which subsequently allows for novel, higher-level computational technologies to exploit the benefits of processed and shared digital shadows (cf. chapter ▶ "Actionable Artificial Intelligence for the Future of Production").

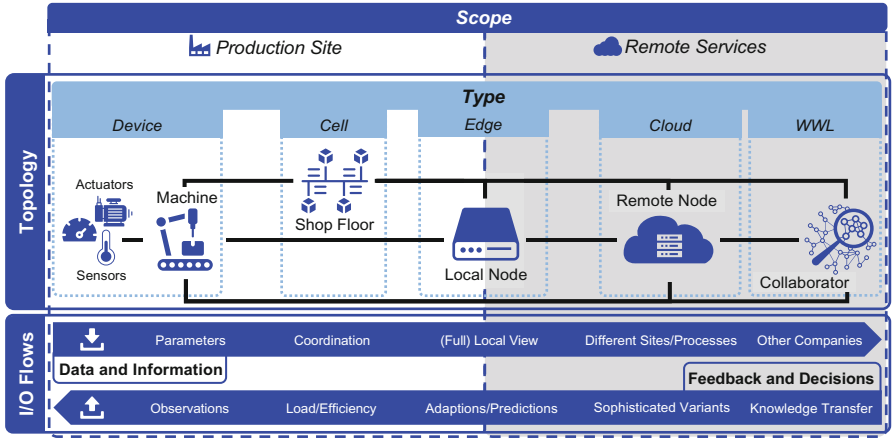

**Fig. 1** The digital industrial *Infrastructure of Production* is characterized by a nonlinear topology, individual data and information flows for all nodes, and a scope increasingly reaching from isolated production sites to remote services and beyond company borders to realize a World Wide Lab

With a specific focus on the underlying, networked infrastructure, we identify several research questions, especially when having real-world deployments in mind:

- How can we design a reliable and scalable *Infrastructure of Production*, i.e., as needed for the IoP with its unique processing and security requirements, that is capable of handling enormous amounts of data (rates) while being able to integrate heterogeneous, distributed models and data streams on time?
- How can we flexibly deploy workloads in such a decentralized network-enabled environment to allow for adaptable and high-performant control decisions?
- Which approaches enable stakeholders to exchange massive and heterogeneous datasets in such environments while considering their confidentiality needs?

To answer these questions, we focus our efforts on five research areas that build the core of an evolved industrial infrastructure. This way, we cover the broadest possible spectrum of hierarchical levels of production (cf. Fig. 1), i.e., we contribute comprehensively to addressing these pressing research questions for real-world use.

In this work, we elaborate on (1) methods for the efficient processing of production data in motion and at rest throughout the node topology illustrated in Fig. 1. In the context of production, data is recorded by sensors, machines, and devices. It is further forwarded via edge servers to the cloud and to any collaborators within the WWL. At times, it is also processed directly while being networked in the infrastructure. In line with Fig. 1, insights gained from the data and decisions made are fed back to the shop floor and machines to decisively influence production processes. Further areas investigate cross-sectional functions required by all hierarchical levels, such as (2) the overarching interoperability of systems, (3) the controlled exchange of data with various stakeholders, (4) the current state of and future improvements for network security, and (5) the enabling of privacy-preserving industrial collaboration. Furthermore, successfully addressing the research questions, i.e., adapting the infrastructure according to the needs of production technology and the IoP, especially with its novel confidentiality and privacy requirements, can only succeed if different disciplines (domain experts) collaborate (Brauner et al. 2022), in particular given the demand for concepts that scale to industry needs. Otherwise, valuable knowledge will remain in stakeholder-specific data silos where production experts cannot access and utilize it, i.e., valuable potentials are lost. Thus, evolving today's industrial landscape, its data processing, and the foundation for collaborations into a digital industrial *Infrastructure of Production* is of utmost importance.

## 2 State of the Art: Challenges for the Infrastructure

Before we describe how we aid the evolution of the infrastructure in Sect. 3, we first provide an overview of and derive challenges for the current *Infrastructure of Production* (Sect. 2.1). Afterward, we discuss the state of the research areas that can, once being addressed, enable the Internet of Production (IoP) (Sect. 2.2).

### 2.1 An Overview of the Infrastructure of Production

Fueled by the IoP, the technical infrastructure underlying industrial production is in a phase of drastic transformation. As shown in Fig. 1 (left), traditionally, industrial devices such as machinery and sensors have only been networked *within* one production site. Typically, a *cell* in a production site corresponds to one shop floor (cf. chapter ▶ "Model-Based Controlling Approaches for Production Processes Using Digital Shadows") or modern production concepts, such as lineless mobile assembly systems (cf. chapter ▶ "Resilient Future Assembly Systems Operation in the Context of the Internet of Production"). Devices within one cell have been interconnected within a dedicated process network with no or only severely limited interconnection to other networks (within the same factory or company), let alone the Internet. With the ongoing digitization, companies tend to move data storage and processing from isolated cells to local nodes at the edge of production, potentially combining data from multiple cells and even across different production sites, but still within and in the control of the same company. Thus, industrial deployments typically still confine knowledge in stakeholder-specific data silos due to omnipresent security and privacy concerns. Consequently, any exchange of data and thus collaboration has not been possible by traditional technical infrastructures underlying industrial systems.

Motivated by the manifold benefits promised by the IoP (Pennekamp et al. 2019a), technical infrastructures of production scenarios increasingly shift toward remote services, including cloud computing services, as highlighted in Fig. 1 (right). This shift is mainly driven by the desire to realize digital shadows for various processes and tasks in production (cf. chapter ▸ "The Internet of Production Digital Shadow Reference Model for Worldwide Production Labs") as well as digital shadow-integrated machine learning and artificial intelligence (cf. chapter ▸ "Actionable Artificial Intelligence for the Future of Production"), demanding access to various kinds of data from multiple sources as well as requiring immense computational and storage resources. To fulfill this demand, storage and processing of various kinds of production data and corresponding digital shadows is increasingly moved outside the sphere of individual production sites, ranging from edge computing (still mostly in control of a single company), over remote nodes in the cloud, to joint processing and storage at and with collaborators in a globally interconnected World Wide Lab (WWL). Moreover, today's deeply rooted confidentiality concerns in industry still prevent the utilization of digital shadows, data lakes, and industrial collaborations as companies and stakeholders understandably call for appropriately secured approaches. Thus, other than previous paradigms without access to sufficient domain expertise, interdisciplinary research on the IoP can directly account for these additional challenges.

To provide a solid foundation on the technical level to realize these functionalities and ultimately capitalize on the various benefits of the *Infrastructure of Production* and the expected impact of digital shadows, different streams of research need to be tackled. More concretely, providing a fundamental technical infrastructure requires further research efforts on (1) adapting data and stream processing to heterogeneous data from distributed sources, (2) ensuring data interoperability between systems and production sites, (3) exchanging and sharing data with different stakeholders, (4) network security approaches addressing the risks of increasing interconnectivity, and (5) cybersecurity architectures to enable secure and privacy-preserving industrial collaboration. Orthogonal to these technical aspects of the infrastructure, human aspects w.r.t. to the people working on and interacting with production processes (cf. chapter ▸ "Human-Centered Work Design for the Internet of Production") as well as requirements resulting from new business models and relationships in the WWL (cf. chapter ▸ "Design Elements of Platform-Based Ecosystem and Industry Applications") need to be considered.

## 2.2 Research Areas for the Infrastructure of Production

Resulting from the identified key challenges for the *Infrastructure of Production*, we now outline five research areas by discussing the corresponding state of the art and associated research questions. Together, these areas make up the center of our envisioned infrastructure. In particular, the *processing of data* and *device interoperability* enable fundamental interactions between the different entities on a device level. *Data security and quality*, as well as *network security*, provide concepts for securing transmitted data and device interactions. Finally, an *infrastructure for industrial collaboration* promises to enable secured interactions on a higher level of abstraction, e.g., across company borders, to enable the exchange of knowledge.

### 2.2.1 Scalable Processing of Data in Motion and at Rest

Data is recorded, transmitted, stored, and processed throughout the whole topology, as illustrated in Fig. 2. Conceptually, it is either in motion, in use, or at rest. Data *in motion* refers to data that moves from a source to a destination within a private or public network. Data *in use* is data that is currently being accessed, processed, or updated. When data is persisted on nonvolatile storage, such as (edge) cloud storage or (industrial) data lakes, it is called data *at rest*. In the following, we separately discuss the subareas of data stream management and analysis and the processing of the data at the network edge, in the cloud, and during transmission.

**Data Stream Management and Analysis** The huge number of devices and sensors in modern industrial manufacturing sites leads to the production of massive amounts of data in the form of continuous unbounded data streams with a high frequency. Data Stream Processor (DSPs) are systems tailored to the management and analysis of data streams and support the efficient querying and implementation of near real-time applications, such as anomaly detection or alerting. However, the data size emitted by sensors and machines is, in many cases, so huge that the transfer to a central system for further processing and analysis would lead to a network overload (Pennekamp et al. 2019a). Hence, for the data stream processing to become

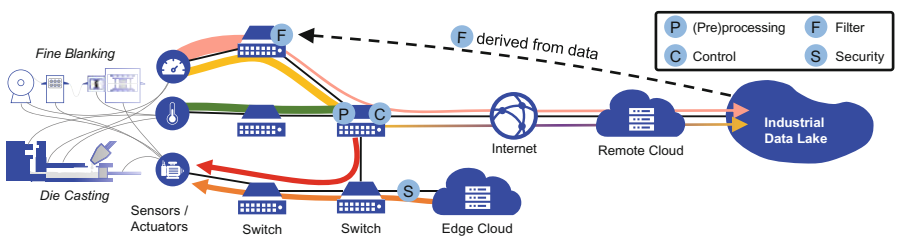

**Fig. 2** Sensor data is transmitted and processed at various locations throughout the *Infrastructure of Production*. Data streams can already be filtered (F) or (pre)processed (P) in the network. Moreover, they can be protected using secure protocols or gateways (S). Switches can generate control commands (C) significantly closer to the production process than other processing entities

horizontally scalable, the workload must be pushed further toward distributed machines, i.e., to the fog and the edge of the network. This approach promises efficient processing where needed while introducing new challenges to DSPs (Zeuch et al. 2020). Here, as illustrated in Fig. 2, one fundamental question is where to place the operators, such as filters (F), of the processing graphs maintained by these DSPs.

Furthermore, such a processing infrastructure must also support highly heterogeneous data from various sources, such as sensors, machines, and cameras. Also, when many components and sensors are involved, data quality can be highly variable, as these components may fail or emit invalid data. Hence, data quality needs to be assessed and considered in each step of the infrastructure. Produced data further needs to be enriched with standardized domain-specific and processual metadata, reflecting the context and provenance of the data stream elements. The metadata increases its value for higher-level applications and makes it suitable for data exchange in the IoP. Different levels of data richness hereby represent digital shadows of different granularities, serving the needs of different applications.

**Edge Computing** An edge device is any type of computational resource that resides between the original data source and cloud-based data centers. Such devices range from very limited industrial components, often with interfaces to industrial control systems, to powerful computing units with special hardware specifications, e.g., additional graphics processing units (Qi and Tao 2019). The goal of edge computing is to transform and reduce data locally by eliminating redundancies as close to the data source as possible with limited resources to provide faster feedback on events in the data or to reduce the data volume load on the network. The IoP presents new challenges for edge computing, as data volume and velocity, the feedback required for control mechanisms, and the complexity of models require a fine-tuned computational hierarchy (Glebke et al. 2019). Data is only available to a limited extent on a local level. Thus, calculations, models, and decisions based on them can only include a slice of the globally available data. The key challenge lies in the use of limited resources and data access to extract information, reduce redundancy, and provide feedback latencies that fulfill use case-specific requirements. For example, when considering the architecture of Fig. 2, an edge cloud could perform computations close to the machine to save latency compared to processing in a remote cloud that first mandates a time-consuming transmission of relevant data.

**Cloud Computing** A big challenge in current manufacturing information systems is the organization of data flows, data collections, and their analysis in a centralized manner, allowing all participants to either ingest, modify, or extract information. To this day, most manufacturing systems use sensor-actuator systems without giving access to outside IT systems (i.e., Operation Technology (OT)-Internet Technology (IT) separation (Garimella 2018)) to allow data analytics and condition monitoring. The usage of manufacturing data across multiple production steps and distributed production sites and along the whole lifecycle holds great potential for saving costs, for a more efficient use of resources, and for an increase in product quality (Rath et al. 2021). However, a centralized cloud infrastructure that integrates all relevant production data is a very complex system that is difficult to build, maintain, and run, as today's data processing systems are characterized by a fragmentation of computing and storage resources. Reasons for this fragmentation are either missing organizational infrastructures like user management across multiple systems or vendor lock-ins which limit the accessibility of data by the use of proprietary technology. Manufacturing sensors and actuators can produce data up to a few GHz, which typically results in enormous amounts of data. As discussed before, the amount of data produced by a manufacturing system puts large requirements on the networking, processing, and storage infrastructure. These data streams only produce data during their use, i.e., the stream of data is more inconsistent and produces large bursts. Having a flexible and agile infrastructure that continuously adapts to these bursts and also provides an automated way of scheduling and scaling applications inside the data center is a large challenge (Brockmann and Kröger 2021). A central concept in the IoP is the data lake (cf. Fig. 2), which represents a repository of all (historical) production data. It is implemented as a scalable distributed system and storage while being controlled centrally through a dedicated control plane.

**(In-)Network Capabilities** The communication network, over which data is transmitted and streamed, is the central connector of all entities within the IoP as it is used to communicate all nonlocal information. In a traditional sense, the network is seen as a "dumb" connection provider that only delivers data. However, research has already illustrated that data rates needed for industrial production processes can become challenging for traditional network setups (Glebke et al. 2019). Additionally, physical signal propagation latencies can also become an influencing factor as soon as computation and/or storage components move to remote facilities, e.g., in the form of cloud providers or the envisioned centralized data lake. Consequently, the existing network infrastructure represents a potential bottleneck for the IoP.

Fueled by novel network programming concepts, such as SDN and P4, research has once again turned its focus to finding ways to leverage previously unused compute resources within the network. This trend is commonly known as *In-Network Computing (INC)* (Sapio et al. 2017). While the exact scope of INC is not yet clearly defined, especially regarding potential overlap with edge computing and whether INC should only refer to computations directly on networking devices, there is already significant work that studies which compute tasks can be best mapped to networking devices and how (Ports and Nelson 2019). In our example in Fig. 2, we could, e.g., place a simple processing function (P), such as aggregation, on a switch to significantly reduce the amount of transmitted data. The main challenges arise from the limited computational complexity supported by such devices as they are designed for high-speed packet processing, but only simple calculations. In addition to the challenge of which calculations should be performed using INC, open questions remain as to how and where functionality should be placed within the network and, more importantly, how the functionality should interact with the existing end-host-focused computation and communication schemes. As of today, a generalized, scalable INC framework that solves all or even most of the mentioned challenges is still missing.

### 2.2.2 Device Interoperability

To enable the automated collection of sensor and measurement data, measuring systems (JCGM 2012) must be integrated into the *Infrastructure of Production*. Due to the wide plethora of distributed and used systems, all having their individual manufacturer- and device-dependent interface, this integration is a nontrivial task and requires manual adaption and integration each time (Bodenbenner et al. 2020). The respective individuality and dependency are essentially expressed in the following three aspects (Montavon et al. 2019): (i) manufacturers choose their favored programming language for implementing the systems logic and the API, (ii) the protocol and format for exchanging data differs from manufacturer to manufacturer and often even from device to device, and (iii) the data and system model, which forms the base of the interface of the system, is usually developed from a physical point of view, instead of a functional one, resulting in very low interoperability. Moreover, due to the trend of coupling the internal sensor logic and the communication interface, highly complex cyber-physical systems are formed, which increase implementation, integration, and maintenance efforts (Thramboulidis and Christoulakis 2016).

To seamlessly integrate measuring systems into the overarching digital industrial infrastructure, maximizing the interoperability of used cyber-physical measuring systems is crucial. Concerning the digital infrastructure, this goal requires solving the aspect of technical and syntactic interoperability (Bodenbenner et al. 2021). Although several interoperable data formats and communication protocols are already in use in industry, an approach is missing that decouples the development of internal sensor logic from the communication interface, i.e., the incorporation of interoperable data formats and communication protocols. Solving that would reduce the effort of integration and maintenance of measuring systems in an industrial infrastructure. The challenges and demands described here are of utmost relevance in modern assembly paradigms, such as lineless mobile assembly (Hüttemann et al. 2019). Consequently, we investigate corresponding use cases in our research.

### 2.2.3 Data Security and Data Quality

Given the sensitive nature of (production) data, decisions and data management plans of stakeholders are frequently driven by concerns about data security to secure their competitive advantages, i.e., companies fear a loss of control or unintentional data leaks (Brauner et al. 2022). As a result, in today's environments, data is mostly retained and encapsulated locally at a company (Gleim et al. 2020), potentially even at a single production site or within a specific production cell. Thus, only a single stakeholder can utilize such data silos, resulting in isolated data across the industry. This situation severely hinders industry-wide process and product improvements.

As data is typically kept locally, companies frequently consider security measures unnecessary and neglect to implement data security and privacy solutions and policies in practice. As a result, companies also tend to favor on-premise computing over cloud computing. However, the expected benefits of (i) increasingly automated decision-making within industry (e.g., using machine learning) and (ii) initiatives (including industrial dataspaces (Geisler et al. 2022)) call for revised data security policies that enable companies to globally utilize all available information.

Especially the real-world impact on production lines mandates suitable approaches that can address the security and safety needs of industry while handling the vast amount of (production) data (Henze 2020). In this context, the development of new concepts for data sovereignty, authenticity, verifiability, and accountability has to be considered. So far, these aspects were mostly out of scope as (i) data was rarely used to manipulate live processes and (ii) data was not shared between stakeholders.

Moreover, when considering the trade-off between privacy and transparency, the reliability of information becomes increasingly important. To allow for an ideal utilization, data should be sensed accurately and in a trustworthy manner (Bader et al. 2021), should be authentic and correct (Pennekamp et al. 2020a), and should be semantically enriched (Gleim et al. 2020). Production data that is available within a semantical framework increases the value of such data for downstream data users by lowering the inefficiencies associated with the data exchange as the domain knowledge among stakeholders generally varies, which results in inefficiencies for both data users and data providers. Thus, through semantic enrichment, a frictionless integration with downstream users is enabled. Consequently, semantically enriched production data is far more valuable than raw data, since it is directly available for efficiency gains (without the need for excessive preprocessing). On a similar note, when proposing novel approaches, legal aspects (e.g., liability questions) must be taken into account. In the past, suitable approaches from computer science were not vetted because production environments were neither digitized nor interconnected.

### 2.2.4 Network Security

Sharing data between stakeholders requires significantly intensified communication between all components and layers. Thus, formerly isolated production networks need to be interconnected with other production networks as well as office networks and the Internet. This development facilitates the risk of eavesdropping attacks on sensitive business information or malicious takeover of production machines. Such attacks can not only lead to monetary loss due to the disclosed business secrets but may even cause production outages or create harm to humans (Brauner et al. 2022). Hence, securing these networks is a key requirement.

Since traditional industrial communication protocols, e.g., Modbus, were designed for communication in isolated environments, their design does not include any security mechanisms (Dahlmanns et al. 2020). Furthermore, especially older embedded industrial devices often lack the computational resources to perform state-of-the-art cryptography operations, which becomes particularly problematic in the face of the upcoming shift toward post-quantum cryptography (Henze 2020). Nevertheless, even today, operators connect industrial devices to the Internet while relying on these insecure traditional protocols (Mirian et al. 2016; Dahlmanns et al. 2022; Nawrocki et al. 2020). Consequently, attackers can access these production devices without restrictions, alter messages, or eavesdrop on exchanged information.

In recent years, traditional protocols were retrofitted with Transport Layer Security (TLS), the state-of-the-art protocol for secure communication on the Web. While these protocol versions provide confidentiality, integrity protection, authentication, and access control, their security *in practice* depends on a regularly updated configuration keeping up with changes in the security landscape, e.g., to account for outdated ciphers or hash functions (Dahlmanns et al. 2022). Hence, operators frequently need to assess and adapt their security configurations accordingly. However, security analyses indicate that 42 % of all TLS-enabled industrial protocol deployments on the Internet, i.e., deployments which are reachable in the IPv4 address space, show security deficits (Dahlmanns et al. 2022). Additionally, although OPC UA, the most promising modern industrial communication protocol, was designed with security in mind, research shows that 92 % of the Internet-reachable OPC UA deployments are configured with security deficits (Dahlmanns et al. 2020).

Besides securing communication, exposing networks to the Internet also requires mechanisms to reliably detect potentially remaining attacks (Henze 2020). However, unique opportunities for detecting advanced attacks in cyber-physical systems such as industrial control systems, e.g., by leveraging semantic or process knowledge, remain typically unused today.

### 2.2.5　Infrastructure for Secure Industrial Collaboration

Primarily due to deeply rooted confidentiality concerns (cf. data security), extensive data sharing between stakeholders has not yet been implemented in industrial practice. Hence, corresponding (secure) information flows, while widely researched, remain mostly untapped so far (Pennekamp et al. 2019b). As a consequence, companies cannot fully benefit from the potential of industrial collaboration. Thus, research has to demonstrate the benefits of industrial collaboration to ease its deployment in production environments through a dedicated infrastructure.

The IoP and the proposed WWL further envision modern, dynamically evolving business relationships to address tomorrow's objectives (costs, quality, sustainability, and others (Pennekamp et al. 2021c)) in production. These short-lived relationships significantly challenge today's established level of trust. Thus, to mitigate these concerns, stakeholders demand (proven) technical security guarantees, which underline a strong protection of their sensitive information at all times. Likewise, as automated adaptations based on external information are envisioned, industrial collaborations can only succeed if they ensure a safe operation of all processes (in terms of both human operators and the environment) (Pennekamp et al. 2019b; Henze 2020), while still yielding added value for participating companies.

When sharing data, companies further expect mechanisms to automatically evaluate any allowed secondary use of their shared data, e.g., using data usage policies (Henze 2020; Henze et al. 2016). Otherwise, their concerns could effectively prevent collaboration in industry. Importantly, these challenges do not only concern stakeholders along supply chains but also across supply chains, e.g., if operators of similar machinery collaborate. Importantly, even direct market competitors could collaborate if they are sufficiently supported through technical building blocks. Yet, solutions are only slowly emerging from the domain of traditional cloud computing.

## 3  Evolving Today's Infrastructure for Future Industry Use

Based on the challenges in Sect. 2.2, subsequently, we describe for each research area which solutions we have proposed so far. Additionally, we point out further directions that we will pursue to establish a capable *Infrastructure of Production*.

### 3.1  Scalable Processing of Data in Motion and at Rest

In the area of data processing, we separately elaborate on our contributions and further research directions concerning the different building blocks of data stream management and analysis as well as edge, cloud, and in-network computing.

**Data Stream Management and Analysis**  Industrial environments and production sites require data management infrastructures that can handle massive amounts of data in a short time. To tackle the problem of network overload when sending all data to a central data stream processor, we work toward a scalable infrastructure that can distribute a continuous query over a multi-level topology of edge, fog, and cloud computing nodes. An abstraction for a continuous query is a directed acyclic graph of operators, which execute, e.g., filter or windowing operations on streams. The distribution of operators over the graph of potentially changing network nodes is a challenging optimization problem (Cardellini et al. 2016). Thus, various dimensions need to be weighed, e.g., latency requirements or hardware capabilities.

We are working toward a dynamic, robust, secure, and smart infrastructure for production that is able to include different kinds of dimensions, and we particularly investigate dimensions that are relevant for these industrial settings, such as data economy, privacy, and data quality. To address the issues of hardware heterogeneity, we are developing a lightweight and system-agnostic operator library along well-known stream semantics to also enable higher-level declarative and procedural query languages. Furthermore, also new methods for multi-query optimization must be investigated which fit the envisioned infrastructure, benefiting data economy.

Streaming data produced in manufacturing processes is not only valuable for real-time applications but also needed for historical analyses spanning a longer time period. Hence, the data needs to be persisted in long-term storage solutions, such as data lakes. Additionally, a crucial aspect for the Internet of Production (IoP) is the sharing of data with other stakeholders in the World Wide Lab (WWL) and providing input to digital shadows (Brauner et al. 2022). For the scenarios above, properly annotating the data is paramount to prepare it for sharing and reuse. To that end, we develop a metadata model and management concept that (i) allows for the general annotation of data streams along multiple categories and dimensions of metadata and (ii) considers domain-specific semantics and vocabulary from manufacturing for the annotation. Depending on where the metadata emerges, the corresponding functions also need to be deployed to the different levels of the streaming infrastructure topology. In addition, based on previous work (Geisler et al. 2016), we strive to integrate intelligent data quality assessment and improvement to detect data flaws early in the processing pipeline and also prevent swamping the permanent storage into the infrastructure from the cloud to the edge nodes.

**Edge Computing** Edge devices provide only limited computing resources that have to be leveraged to reduce the amount of data and condense the information density. Raw digital representations of manufacturing processes are often based on multivariate time series that contain a variety of overlapping physical effects encoded in the signals. To realize the potential of this hidden data, domain experts who understand the process and its effects must prepare the data appropriately and make it available for further analysis. The incorporation of domain knowledge at the edge is crucial, since the sooner the data is preprocessed and labeled in a domain-specific way, the less redundant or irrelevant information needs to be passed on. Note, however, that a purely human-based method of preprocessing for knowledge may neglect important unknown effects (Liewald et al. 2022). Thus, edge-based preprocessing pipelines need to incorporate domain knowledge-based approaches into data-driven approaches to fully leverage the power of edge systems.

The processing of a specific manufacturing process in sheet metal forming utilizing high-frequency sensor systems ranging from 10 kHz to 10 MHz reveals the need for edge computing to reduce the data load (Glebke et al. 2019). In this study, nine sets of time series data represent the collective data load during the manufacturing process, which can be broken down into process phases. Each process phase is characterized by specific physical effects, so that only a small part of the time series is useful to include in the modeling. Leveraging this fact, the amount of data relating to a forming operation can be reduced by up to 70 % if, e.g., only the wear or another specific characteristic is of interest (Niemietz et al. 2020). Further studies show that information in raw sensor time series is often redundant, yet classical feature engineering methods based on domain knowledge only partially work (Niemietz et al. 2022). Similarly, features selected by experts are highly redundant and unsuitable for further compression (Unterberg et al. 2021). However, traditional dimension reduction methods can further considerably reduce the amount of data (Bergs et al. 2020). A domain knowledge-based approach combined with methods that independently learn features of time series data can already yield good monitoring capabilities utilizing only computational edge feasible models (Niemietz et al. 2021).

For sheet metal forming and fine-blanking in particular, the IoP enables dynamic processing of all available information specifically tailored to use case needs. Thus, data can be processed locally and only transmitted to the cloud if needed. Otherwise, sensed data, and valuable information within, is lost as the amount of data quickly surpasses the network capabilities or it violates latency requirements in control and feedback loops. These approaches show that using only computationally feasible models for data reduction at the edge in combination with domain knowledge considerably reduces the amount of data that needs to be transmitted to (cloud-based) data centers. However, a comprehensive framework for data reduction and condensation of industrial processes considering the locality of data, models, and knowledge is still missing, as well as methods for combining knowledge and raw data at the edge.

**Cloud Computing** Cloud technologies, either provided by commercial cloud providers or custom on-premise systems, offer the possibility of dynamically adapting required resources according to demand. This scalability applies to all areas of the data infrastructure: computing power, memory, networking bandwidth, and storage. In addition, clouds are the ideal medium for creating systems that have a uniform level of abstraction and thus appear as a single infrastructure that is available to technical as well as human producers and consumers of data, regardless of their physical location. In this sense, cloud technologies provide not only the computing resources needed but also provide ways to control and manage these systems through the so-called control plane (Casquero et al. 2019). Therefore, cloud infrastructure reduces the complexity of running applications in a distributed environment and abstracts away implementation details of components (Beyer et al. 2016).

Ideally, a cloud infrastructure also provides a set of additional services that centralize functionalities shared by multiple components, such as user management, condition monitoring, or network security. As part of our research, a prototypical Kubernetes-based manufacturing data and control hub has been developed, which greatly reduces the effort of scheduling and hosting support applications for manufacturing. This computing cluster lays the foundation for automated data streaming of machine sensor data and their analysis by providing computing, storage, and connectivity resources on demand. The system was designed in a way that the ingestion of data streams, the persistent storage of data in databases or data lakes (Rudack et al. 2022), its automated analysis, firewall rules, user rights, and analytics software are all managed by the Kubernetes operator pattern (Verma et al. 2015), which greatly lowers the cognitive load caused by running and maintaining such a system. Even software-controlled manufacturing machines have been managed by this control plane which removes the barrier between Operation Technology (OT) and Internet Technology (IT) completely, since the manufacturing machine's software is deployed, run, maintained, and monitored in the same centralized manner as a cloud-based database or support application. Therefore, the cluster acts as a centralized connector which enables the usage of production data along the complete lifecycle of the production process. The use of open-source software not only avoids vendor lock-ins but also guarantees interoperability between systems. Through containerized software applications and by following established cloud-native principles, our prototype is transferable to any cloud provider. To assess the viability of our infrastructure in production, we connected a 500 t horizontal high-pressure die casting machine and its auxiliary cell systems to a prototypical data lake. This design allows us to experiment with a full-scale physical machine on premises. In particular, we utilize a testbed that includes complex production machinery with multiple PLCs, an edge server, and a dedicated cloud infrastructure (data lake) (Rudack et al. 2022).

**(In-)Network Capabilities** Today's communication systems follow a strict interpretation of the end-to-end principle (Saltzer et al. 1984), i.e., the network only delivers packets without modifying them (Kunze et al. 2021c). Thus, In-Network Computing (INC) is typically not supported out of the box. Aggravatingly, the exact scope of INC remains undefined, and identifying which domains can benefit most is still an ongoing process (Kunze et al. 2022). Research already investigates how INC can be included in existing communication infrastructures (Kunze et al. 2021c) and what kind of functionality can be provided by INC (Kunze et al. 2022).

The main envisioned benefits of INC for industrial processes are higher data rates and lower latencies as networking devices can process packets at line rate and are located close to the processes. The straightforward latency benefits have already been studied and demonstrated by related work (Cesen et al. 2020), mainly focusing on In-Network Control (cf. chapters ▶ "Model-Based Controlling Approaches for Production Processes Using Digital Shadows, Resilient Future Assembly Systems Operation in the Context of the Internet of Production"). In our work, we investigate which functionality can be enabled by INC in spite of specific hardware constraints.

In modern industrial scenarios, all entities (robots, machinery, etc.) are tracked using various metrology systems, such as iGPS or laser trackers (cf. chapter ▶ "Resilient Future Assembly Systems Operation in the Context of the Internet of Production"). These systems use different formats to capture locations, e.g., using different coordinate systems, and metrology information thus needs to be transformed into a common scheme. In one of our works, we investigate how well the required coordinate transformations can be deployed on programmable networking hardware (Kunze et al. 2021a). While we find that there are indeed challenges requiring heavy workarounds, we also see that INC can achieve low latencies and high accuracy, as well as significantly higher packet processing rates than end-host-based applications.

We further investigate data preprocessing using INC. More specifically, we intend to deploy an INC platform to react to different process phases and dynamically adapt how data is preprocessed and where it is forwarded (Kunze et al. 2021b). In this approach, the INC platform summarizes sensor information in well-defined intervals and uses a local clustering algorithm to distinguish process phases. We offload heavier analysis functionality to a slower control plane which can then define which actions are supposed to be taken depending on the currently identified process phase.

Overall, these approaches showcase that INC can be sensibly used in industrial contexts. However, there are still many remaining questions, e.g., regarding the inclusion of such approaches into existing communication infrastructures (Kunze et al. 2021c), that need to be solved before INC can be widely deployed.

### 3.2 Device Interoperability

To make measuring systems technically interoperable, we aim toward a shift in the role of measuring systems. Instead of considering a measuring system as an integral component of a production process of industrial application, we interpret measuring systems as independent micro-services that offer their services, i.e., the acquisition and provision of measurement values, to different applications and stakeholders. This idea results in the concept of Cyber-Physical Measuring System (CPMSs), creating the basis of modern infrastructures in industry.

We call such a system a FAIR sensor service (Bodenbenner et al. 2021) which is mainly defined by the following three attributes: (i) The CPMS must advertise its offered service and ensure high quality of the provided measurement data, i.e., the data (and the service itself) must conform to the FAIR principles (Wilkinson et al. 2016). (ii) The CPMS senses a change in a physical quantity, quantifies this change as a relative or absolute measurement value of that quantity, and digitizes this value. (iii) For the environment, an interoperable interface characterizes the sensor, which mainly consists of a functional information model and a standardized, manufacturer-independent, and device-agnostic data format and communication protocol.

With that, the interoperability of the measuring device is increased, and the heterogeneity of communication interfaces and data formats is significantly reduced. However, joining those three concerns into one CPMS also increases complexity and requires expertise in three different domains: (i) implementation of the internal sensor logic, (ii) defining a FAIR (meta)data model, and (iii) the development of the communication interface. To decouple these three aspects, we propose a novel three-layer architecture for FAIR sensor services (Bodenbenner et al. 2021) and utilize model-based software development to leverage generalizable parts, as, e.g., the communication interface, by developing a domain-specific modeling language called SensOr Interfacing Language (SOIL) (Bodenbenner et al. 2020). Based on a simple meta-model, the data model of the interface of the FAIR sensor service can be developed without any knowledge of communication protocols and data formats by the developer of the measuring device. Based on the interface description written in SOIL, a template for implementing the internal sensor logic is generated in a general-purpose language (e.g., Python or C++), to which the non-generalizable implementation of the internal sensor logic is injected manually. Furthermore, a RESTful HTTP server and an MQTT publisher are generated, such that there is no additional effort required regarding the connectivity of the measuring device.

To fully realize FAIR sensor services, we are currently researching the automatic generation of metadata schemata from SOIL models, as well as the inclusion of more target languages, communication interfaces, and data formats, which will eventually result in fully, semantically interoperable interfaces as we illustrate in Fig. 3.

### 3.3    Data Security and Data Quality

Enriching existing (industrial) infrastructures with appropriate data security requires developments along multiple dimensions. As the most basic mechanism, data usage policies (Henze et al. 2016), which efficiently formulate allowed utilization of sensitive data, allow stakeholders to express their privacy and data sovereignty needs. For example, companies can specify details about the suitability of using

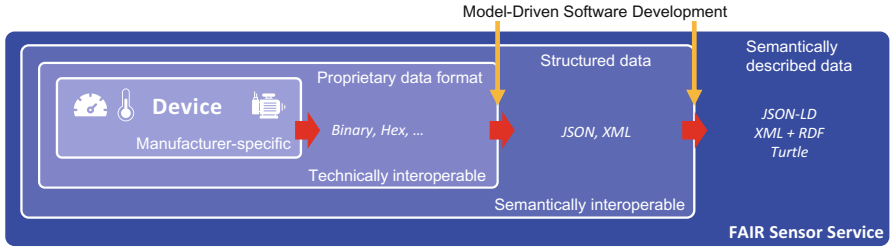

**Fig. 3** Our proposed interoperability hierarchy of interfaces for Cyber-Physical Measuring System (CPMSs). While manufacturer-specific interfaces are mostly non-interoperable, we can achieve interoperability by providing context information through semantically annotated (meta)data

specific cloud infrastructures (Henze et al. 2020). However, these policies require an underlying physical infrastructure, e.g., a cloud storage system, that is also capable of enforcing them (Henze et al. 2020). Moreover, recently proposed dataspaces still need to work on providing corresponding technical guarantees (Lohmöller et al. 2022).

Thus, complementing the efforts of developing and enforcing data usage policies, we also pursue the research direction to provide technical security guarantees to companies. In particular, to evolve today's infrastructures, we rely on well-known secure building blocks (Pennekamp et al. 2019b) from the area of privacy-preserving computing to implement data security. For example, we rely on attribute-based encryption to realize reliable, yet dynamic access control within supply chains (Pennekamp et al. 2020b; Bader et al. 2021). Similarly, companies can utilize homomorphic encryption (which enables computations on encrypted data) to protect their sensitive information when performing computations on joint data, e.g., in the context of performance benchmarking, without significantly decreasing utility in practice (Pennekamp et al. 2020d). Thus, the presented and related building blocks are well-suited to implement secure offloading (e.g., to cloud environments) in industry.

Moving toward the challenge of ensuring data quality, we propose different mechanisms to improve the accountability of participating companies. First, we employ blockchain technology to establish technical trust anchors (Pennekamp et al. 2020b; Bader et al. 2021; Wagner et al. 2022b). These trust anchors immutably persist data fingerprints to ensure that the covered information is long-term verifiable. They further ensure that companies can be held accountable, e.g., if faults occur during usage of a manufactured product. We validated our approach using supply chains of cars that involved fine-blanked components. Second, when looking at the exchange of information, our novel digital transmission contracts enable companies to prove that a data exchange took place (Mangel et al. 2021), i.e., companies cannot deny their participation at a later point in time. These mechanisms will improve the reliability of shared information as companies must otherwise fear being held accountable.

Finally, when looking at deployed devices at the production site, we explore the potential of utilizing trusted sensors (i.e., sensors equipped with trusted execution environments) to secure the data collection (sensing) in industry (Pennekamp et al. 2020a). Here, the main advantage of our work is a significant improvement in the correctness and authenticity of processed data, which allows companies to secure the full data lifecycle of their products for the first time.

### 3.4 Network Security

To impede cyberattacks against production deployments and thus prevent production outages and harm to humans, implementing strong network security measures is essential. Here, the evolution needs to be threefold: (i) security must evolve with industrial communication use cases, (ii) operators need support in configuring their deployments securely, and (iii) remaining attacks must be detected.

The interconnection of production deployments leads to novel forms of communication which need to be secured. For example, end-to-end secure communication was rather challenging in the novel publish/subscribe paradigm for industrial communication. With our approach ENTRUST, which transparently enables end-to-end secure communication and integrates seamlessly into existing infrastructures, we allow future publish/subscribe deployments to communicate end-to-end secure (Dahlmanns et al. 2021). Likewise, the unique characteristics of industrial settings and especially the resource constraints of industrial devices require adapting traditional security paradigms, especially in wireless settings. Our research work, e.g., allows us to efficiently realize message authentication for short messages (Wagner et al. 2022a) or to speed up the computation of message authentication tags in the latency-critical input-dependent part through bitwise pre-computations (Wagner et al. 2022c).

From a different perspective, the increase in security mechanisms and protocols equally makes the security configuration of deployments more complex and thus challenging. Hence, operators need support in regularly assessing their current security configuration with regard to the current security landscape and require assistance in configuring their industrial deployments securely. As most of today's assessment tools lack support for modern industrial protocols, e.g., OPC UA, we designed and developed an open-source plugin for the state-of-the-art pentesting software Metasploit that supports operators in analyzing the security of their deployments (Roepert et al. 2020). To support operators in realizing secure configurations, e.g., when a security assessment identifies deficits, up-to-date configuration templates can assist operators by making best practices easily and actionably accessible (Dahlmanns et al. 2022). However, our research shows that such templates should never include any example credentials, since operators often unknowingly do not exchange them and consequently severely weaken their own security (Dahlmanns et al. 2022).

Besides all preventive security measures, intrusion detection and prevention systems are needed to thwart any remaining, especially unknown, attack vectors.

While recent research provides sophisticated approaches leveraging semantic and process knowledge, these approaches rarely find their way into practice, mainly due to their tight coupling to distinct industrial communication protocols and individual datasets. By leveraging commonalities found in industrial communication, our research lays a common ground for realizing widely applicable industrial intrusion detection systems (Wolsing et al. 2022). Furthermore, state-of-the-art industrial intrusion detection systems typically rely on machine learning to detect anomalous behavior. While these systems achieve, in theory, extremely high detection accuracy, in practice, they often miss unknown attacks. To overcome this false sense of security, our novel evaluation methodology assesses whether industrial intrusion detection systems are indeed able to detect attacks they have not been trained on and identifies significant room for improvement to realize efficient and effective machine learning-based industrial intrusion detection systems in practice (Kus et al. 2022).

### 3.5 Infrastructure for Secure Industrial Collaboration

When evolving the production landscape from localized production sites to a globally connected WWL that directly influences decision-making, significant research efforts are needed to realize an infrastructure for industrial collaboration in a secure and privacy-preserving manner. The underlying goal of industrial collaboration is to unlock all available data sources from different stakeholders in real time (Pennekamp et al. 2021b). To implement such a disruptive change within the industrial infrastructure, we envision the gradual implementation of different applications (with increasing levels of automation), as we illustrate in Fig. 4.

First developments will tackle use cases without automated process adaptation, i.e., initially, the infrastructure must provide means to compare information between stakeholders without directly interfering with running processes (Fig. 4, left). Here, a goal can be to provide companies with insights into unrealized potentials

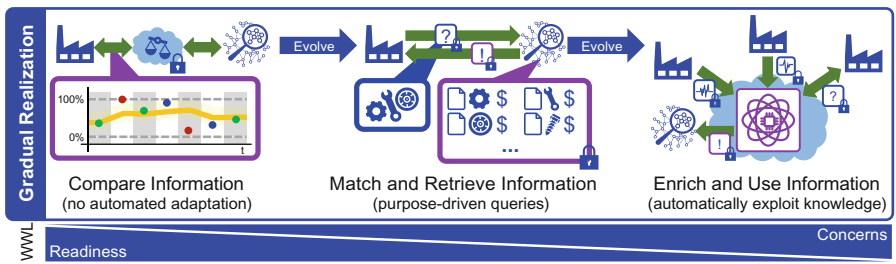

**Fig. 4** Industrial collaborations will be realized gradually. An increasing level of automation requires new approaches and building blocks. Simultaneously, the evolution of the infrastructure will increase the stakeholders' concerns as the sourced technology is not yet proven and tested

by comparing their business performance or process decisions. We exemplarily realized such an architecture using homomorphic encryption (i.e., computations on encrypted data) to preserve confidentiality while performing comparisons alongside various key performance indicators (Pennekamp et al. 2020d). As another application, we are currently pursuing a new secure and privacy-preserving concept to support companies during the establishment and identification of new business partners to account for the dynamic nature of the IoP and the increasing necessity of quickly reacting to customer change requests (Pennekamp et al. 2021a).

Subsequently, moving toward an infrastructure that allows companies to directly improve their local processes by incorporating external information, i.e., by effectively tearing down today's established data silos, requires significant changes to the information flows in industry. To demonstrate the adequate and purpose-driven matching of information, we designed a platform that provides companies with an approach to exchange production process parameters while keeping their sensitive information and queries private (Pennekamp et al. 2020c). Using secure building blocks (oblivious transfers, private set intersection, and Bloom filters), we can keep all information private until an explicit match has been made by a querying company. In general, purpose-driven queries need to be matched with data using semantic information to allow for meaningful process improvements (Fig. 4, center).

Moving forward, the physical infrastructure underlying the IoP is envisioned to evolve into a direction that allows companies to directly feed new insights from external information into their processes (Fig. 4, right). In this context, (cloud-based) federated yet privacy-preserving machine learning constitutes a key research direction that promises direct implications on industrial control loops or adapted production planning. Consequently, these novel information flows must be supported by the underlying physical infrastructure to enable these new collaborations.

## 4 Conclusion

The underlying infrastructure from the sensor to the cloud serves as the foundation of and key enabler for the Internet of Production (IoP) and facilitates the use of its internal multilayered components, such as digital shadows, data lakes, and the World Wide Lab (WWL). The transition from sparsely connected production machines, cells, and sites with mostly isolated and incompatible computing nodes and data silos to an interconnected IoP requires a fundamental redesign and evolution of the underlying infrastructure. We identified open challenges in five research areas for the *Infrastructure of Production* that prevent the realization of the IoP with its distinct requirements for the underlying infrastructure in practice. These issues include limitations rooted in today's security and network architectures across production sites as well as challenges related to the involved computing paradigms, such as edge computing, in-network computing, and the overarching cloud computing. Further research is needed on how to facilitate the secure interaction of network devices, machines, and sensors to enable low-friction collaborations between different stakeholders. All of these challenges must be solved to finally establish the IoP since an adaptable, interoperable infrastructure serves as the enabler for the construction of digital shadows, their exploitation via downstream applications, and a thriving WWL. Related industrial paradigms and research without a connection to the industrial domain, both largely being without the required access to domain expertise, cannot make up for the outlined shortcomings and research gaps.

In this work, we discussed our research agenda and goals on a more general level. However, some key aspects of previous and ongoing research especially exemplify our contribution to the evolution of the industrial landscape and, in particular, the IoP with its unique processing and security requirements. Additionally, we can source from a plethora of realistic real-world use cases, which allow us to approach research challenges in a goal-oriented manner. Our work on coordinate transformations, e.g., demonstrates that novel networking paradigms can be effective solutions for the specific challenges of metrology systems (Kunze et al. 2021a). Moreover, we illustrated how to utilize generative software development to reduce the integration effort of measuring systems by generating interoperable communication interfaces based on a unified data model (Bodenbenner et al. 2020). To further address data security requirements in situations with dynamic business relationships (as envisioned in the WWL), we developed a design that enables companies to share product and production information flexibly and securely (Pennekamp et al. 2020b). Our direct ties to industry even allowed us to evaluate this design using a real-world use case covering an electric vehicle production (Bader et al. 2021). Derived from requirements for secure communication inside envisioned future production plants, we proposed ENTRUST, a novel solution to transparently end-to-end secure publish/subscribe communication (Dahlmanns et al. 2021). Finally, our interdisciplinary research environment allows us to pursue visionary and possibly disruptive ideas. One prime example is our developed parameter exchange as it securely realizes flows of information that currently do not exist in industry due to confidentiality concerns. Such efforts underline the value of research into an *Infrastructure of Production*.

Building on these first advances, we specifically regard our planned in-network process phase detection with its subsequent adaptable data preprocessing (Kunze et al. 2021b) as a promising next step. We intend to show how to directly transfer the latest advances in networking to use in industry, despite the limitation that such initial concepts do not yet exploit the IoP's full potential. In the future, we strive to continually evolve our outlined research areas to advance the general evolution of industrial infrastructures to effectively transform them into a securely and globally interconnected *Infrastructure of Production* with access to a prosperous WWL.

**Acknowledgments** Funded by the Deutsche Forschungsgemeinschaft (DFG, German Research Foundation) under Germany's Excellence Strategy – EXC-2023 Internet of Production – 390621612.

# References

Bader L, Pennekamp J et al (2021) Blockchain-based privacy preservation for supply chains supporting lightweight multi-hop information accountability. Inf Process Manag 58(3):102529

Bergs T, Niemietz P et al (2020) Punch-to-punch variations in stamping processes. In: Proceedings of 2020 IEEE 18th world symposium on applied machine intelligence and informatics (SAMI'20). IEEE

Beyer B, Jones C et al (2016) Site reliability engineering: how Google runs production systems. O'Reilly

Bodenbenner M, Sanders MP et al (2020) Domain-specific language for sensors in the internet of production. In: Proceedings of 10th congress of the German academic association for production technology (WGP'20), vol 20. p100206, Springer

Bodenbenner M, Montavon B et al (2021) FAIR sensor services – towards sustainable sensor data management. Measur Sens 18

Brauner P, Dalibor M et al (2022) A computer science perspective on digital transformation in production. ACM Trans Internet Things 15(2):1–32

Brockmann M, Kröger M (2021) Entering the world of internet of production – get rid of the not-invented-here-syndrome and start making software components foolproof! Edge Native Podcast – 10

Cardellini V, Grassi V et al (2016) Optimal operator placement for distributed stream processing applications. In: Proceedings of 10th ACM international conference on distributed and event-based systems (DEBS'16). ACM

Casquero O, Armentia A et al (2019) Distributed scheduling in Kubernetes based on MAS for Fog-in-the-loop applications. In: Proceedings of 2019 24th IEEE international conference on emerging technologies and factory automation (ETFA'19). IEEE

Cesen FER, Csikor L et al (2020) Towards low latency industrial robot control in programmable data planes. In: Proceedings of 2020 6th IEEE conference on network softwarization (NetSoft'20). IEEE

Chang H, Hari A et al (2014) Bringing the cloud to the edge. In: Proceedings of 2014 IEEE conference on computer communications workshops (INFOCOM WKSHPS'14). IEEE

Dahlmanns M, Lohmöller J et al (2020) Easing the conscience with OPC UA: an internet-wide study on insecure deployments. In: Proceedings of ACM internet measurement conference (IMC'20). ACM

Dahlmanns M, Pennekamp J et al (2021) Transparent end-to-end security for publish/subscribe communication in cyber-physical systems. In: Proceedings of 1st ACM workshop on secure and trustworthy cyber-physical systems (SaT-CPS'21). ACM

Dahlmanns M, Lohmöller J et al (2022) Missed opportunities: measuring the untapped TLS support in the industrial internet of things. In: Proceedings of 17th ACM ASIA conference on computer and communications security (ASIACCS'22). ACM

Garimella PK (2018) IT-OT integration challenges in utilities. In: Proceedings of 2018 IEEE 3rd international conference on computing, communication and security (ICCCS'18). IEEE

Geisler S, Quix C et al (2016) Ontology-based data quality management for data streams. J Data Inf Qual 18(4):1–34

Geisler S, Vidal ME et al (2022) Knowledge-driven data ecosystems toward data transparency. J Data Inf Qual 3(1):1–12

Gelhaar J, Groß T et al (2021) A taxonomy for data ecosystems. In: Proceedings of 54th Hawaii international conference on system sciences (HICSS'21). AIS

Glebke R, Henze M et al (2019) A case for integrated data processing in large-scale cyber-physical systems. In: Proceedings of 52nd Hawaii international conference on system sciences (HICSS'19). AIS

Gleim L, Pennekamp J et al (2020) FactDAG: formalizing data interoperability in an internet of production. IEEE Internet Things J 7(4):3243–3253

Henze M (2020) The quest for secure and privacy-preserving cloud-based industrial cooperation. In: Proceedings of 2020 IEEE conference on communications and network security (CNS'20). IEEE 10(3):1661–1674

Henze M, Hiller J et al (2016) CPPL: compact privacy policy language. In: Proceedings of 2016 ACM on workshop on privacy in the electronic society (WPES'16). ACM

Henze M, Matzutt R et al (2020) Complying with data handling requirements in cloud storage systems. IEEE Trans Cloud Comput 1661–1674

Hüttemann G, Buckhorst AF et al (2019) Modelling and assessing line-less mobile assembly systems. Proc CIRP 81:724–729

JCGM (2012) International vocabulary of metrology – basic and general concepts and associated terms (VIM). OIML V 2-200:2012

Kunze I, Glebke R et al (2021a) Investigating the applicability of in-network computing to industrial scenarios. In: Proceedings of 2021 4th IEEE international conference on industrial cyber-physical systems (ICPS'21). IEEE

Kunze I, Niemietz P et al (2021b) Detecting out-of-control sensor signals in sheet metal forming using in-network computing. In: Proceedings of 2021 IEEE 30th international symposium on industrial electronics (ISIE'21). IEEE

Kunze I, Wehrle K et al (2021c) Transport protocol issues of in-network computing systems. Internet-draft, IRTF. https://datatracker.ietf.org/doc/html/draft-kunze-coinrg-transport-issues-05, work in Progress

Kunze I, Wehrle K et al (2022) Use cases for in-network computing. Internet-draft, IRTF. https://datatracker.ietf.org/doc/html/draft-irtf-coinrg-use-cases-02, work in Progress

Kus D, Wagner E et al (2022) A false sense of security? Revisiting the state of machine learning-based industrial intrusion detection. In: Proceedings of 8th ACM cyber-physical system security workshop (CPSS'22). ACM

Liewald M, Bergs T et al (2022) Perspectives on data-driven models and its potentials in metal forming and blanking technologies. Prod Eng 16:607–625

Lohmöller J, Pennekamp J et al (2022) On the need for strong sovereignty in data ecosystems. In: Proceedings of 1st international workshop on data ecosystems (DEco'22). CEUR Workshop Proceedings

Mangel S, Gleim L et al (2021) Data reliability and trustworthiness through digital transmission contracts. In: Proceedings of 18th extended semantic web conference (ESWC'21), vol 12731. Springer, 265–283

Mirian A, Ma Z et al (2016) An internet-wide view of ICS devices. In: Proceedings of 2016 14th annual conference on privacy, security and trust (PST'16). IEEE

Montavon B, Peterek M et al (2019) Model-based interfacing of large-scale metrology instruments. SPIE 11059

Nawrocki M, Schmidt TC et al (2020) Uncovering vulnerable industrial control systems from the internet core. In: Proceedings of 2020 IEEE/IFIP network operations and management symposium (NOMS'20). IEEE

Neumann WP, Winkelhaus S et al (2021) Industry 4.0 and the human factor – a systems framework and analysis methodology for successful development. Int J Prod Econ 233:107992

Niemietz P, Pennekamp J et al (2020) Stamping process modelling in an internet of production. Proc Manuf 49

Niemietz P, Unterberg M et al (2021) Autoencoder based wear assessment in sheet metal forming. In: IOP conference series: materials science and engineering, vol 1157. IOP Publishing 012082

Niemietz P, Kornely MJK et al (2022) Relating wear stages in sheet metal forming based on short- and long-term force signal variations. J Intell Manuf 33:2143–2155

Pennekamp J, Glebke R et al (2019a) Towards an infrastructure enabling the internet of production. In: Proceedings of 2019 IEEE international conference on industrial cyber physical systems (ICPS'19). IEEE

Pennekamp J, Henze M et al (2019b) Dataflow challenges in an *Internet* of production: a security & privacy perspective. In: Proceedings of ACM workshop on cyber-physical systems security & privacy (CPS-SPC'19). ACM

Pennekamp J, Alder F et al (2020a) Secure end-to-end sensing in supply chains. In: Proceedings of 2020 IEEE conference on communications and network security (CNS'20). IEEE

Pennekamp J, Bader L et al (2020b) Private multi-hop accountability for supply chains. In: Proceedings of 2020 IEEE international conference on communications workshops (ICC Workshops'20). IEEE

Pennekamp J, Buchholz E et al (2020c) Privacy-preserving production process parameter exchange. In: Proceedings of 36th annual computer security applications conference (ACSAC'20). ACM

Pennekamp J, Sapel P et al (2020d) Revisiting the privacy needs of real-world applicable company benchmarking. In: Proceedings of 8th workshop on encrypted computing & applied homomorphic cryptography (WAHC'20). HomomorphicEncryption.org

Pennekamp J, Fuhrmann F et al (2021a) Confidential computing-induced privacy benefits for the bootstrapping of new business relationships. Technical Report, RWTH-2021-09499, RWTH Aachen University

Pennekamp J, Henze M et al (2021b) Unlocking secure industrial collaborations through privacy-preserving computation. ERCIM News 126:24–25

Pennekamp J, Matzutt R et al (2021c) The road to accountable and dependable manufacturing. Autom 2(3):202–219

Ports DRK, Nelson J (2019) When should the network be the computer? In: Proceedings of 17th workshop on hot topics in operating systems (HotOS'19). ACM

Qi Q, Tao F (2019) A smart manufacturing service system based on edge computing, fog computing, and cloud computing. IEEE Access 7:86769–86777

Rath M, Gannouni A et al (2021) Digitizing a distributed textile production process using industrial internet of things: a use-case. In: Proceedings of 2021 4th IEEE international conference on industrial cyber-physical systems (ICPS'21). IEEE

Roepert L, Dahlmanns M et al (2020) Assessing the security of OPC UA deployments. In: Proceedings of 1st ITG workshop on IT security (ITSec'20). University of Tübingen

Rudack M, Rath M et al (2022) Towards a data lake for high pressure die casting. Metals 12(2):349

Saltzer JH, Reed DP et al (1984) End-to-end arguments in system design. ACM Trans Comput Syst 2(4):277–288

Sapio A, Abdelaziz I et al (2017) In-network computation is a dumb idea whose time has come. In: Proceedings of 16th ACM workshop on hot topics in networks (HotNets'17). ACM

Thramboulidis K, Christoulakis F (2016) UML4IoT—a UML-based approach to exploit IoT in cyber-physical manufacturing systems. Comput Ind 82:259–272

Unterberg M, Voigts H et al (2021) Wear monitoring in fine blanking processes using feature based analysis of acoustic emission signals. Proc CIRP 104:164–169

Verma A, Pedrosa L et al (2015) Large-scale cluster management at Google with borg. In: Proceedings of 10th European conference on computer systems (EuroSys'15). ACM

Wagner E, Bauer J et al (2022a) Take a bite of the reality sandwich: revisiting the security of progressive message authentication codes. In: Proceedings of 15th ACM conference on security and privacy in wireless and mobile networks (WiSec'22). ACM

Wagner E, Matzutt R et al (2022b) Scalable and privacy-focused company-centric supply chain management. In: Proceedings of 2022 IEEE international conference on blockchain and cryptocurrency (ICBC'22). IEEE

Wagner E, Serror M et al (2022c) BP-MAC: fast authentication for short messages. In: Proceedings of 15th ACM conference on security and privacy in wireless and mobile networks (WiSec'22). ACM

Wilkinson MD, Dumontier M et al (2016) The FAIR guiding principles for scientific data management and stewardship. Sci Data 3:160018

Wolsing K, Wagner E et al (2022) IPAL: breaking up silos of protocol-dependent and domain-specific industrial intrusion detection systems. In: Proceedings of 25th international symposium on research in attacks, intrusions and defenses (RAID'22). ACM

Zeuch S, Chaudhary A et al (2020) The NebulaStream platform: data and application management for the internet of things. In: Proceedings of 10th Biennial conference on innovative data systems research (CIDR'20). www.cidrdb.org